\documentclass[11pt]{article}

\title{Deterministic Construction of  RIP Matrices in Compressed Sensing from Constant Weight Codes}

\author{Liqing Xu and Hao Chen
  \thanks{L. Xu and H. Chen are with the Department of Mathematics, School of Sciences, Hangzhou Dianzi University, Hangzhou  310018, Zhejiang Province, China. This research  was supported by NSFC Grant 11371138.}}

\begin{document}

\maketitle

\begin{abstract}

\end{abstract}

The expicit restricted isometry property (RIP) measurement  matrices are needed  in practical application of compressed sensing in signal processing. RIP matrices from Reed-Solomon codes, BCH codes, orthogonal codes,  expander graphs have been proposed and analysised. On the other hand binary constant weight codes have been studied for many years and many optimal or near-optimal small weight and ditance constant weight codes have been determined. In this paper we propose  a new  deterministic  construction of RIP measurement matrices in  compressed sensing  from  binary  and ternary contant weight codes. The sparse orders and the number of budged rows in  the  new constant-weight-code-based RIP matrices can be arbitrary. These contant-weight-code based RIP matrices have better parameters compared with the DeVore RIP matrices when the sizes are small.\\



\section{Introduction}

Compressed sensing was proposed by E. J. Cand\'es, J. Romberg, T. Tao, and D. Donoho in 2006 for efficient sampling of sparse signals and has been found vast applications in signal processing. It  is a technique that recovers a sparse signal  ${\bf x} \in {\bf R}^N$ from the measurement ${\bf y}=\Phi \cdot {\bf x}$ which has $k$ non-zero coordinates at sampling rates that are substantially much lower than the Nyquits-Shannon rate (\cite{CRT,Candes,CT,TC,Donoho,Candes1,Candes2}). For  $n$ budged rows and sampling $k$-sparse signals ${\bf x}$ in ${\bf R}^N$ ($N$ columns, $n<N$),  the measurement  is ${\bf y}={\bf \Phi} \cdot {\bf x}$, where $x \in {\bf R}^N$ is a vector with only at most $k$ non-zero coordinates, $\Phi$ is a size $n \times N$ measurement matrix. Effective algorithms to recover ${\bf x}$ from ${\bf y}$ have been proposed in  \cite{CRT,Candes1,Donoho}. We say  that a matrix ${\bf \Phi}$ satisfies the restricted isometry property (RIP) of order $k$ with the constant $\delta_k$ satisfying $0 \leq \delta_k <1$, if for every $k$-sparse vector (that is, only $k$ coordinates of the signal are non-zero) ${\bf x} \in {\bf R}^N$, $1-\delta_k \leq \frac{||{\bf \Phi} \cdot {\bf x}||}{||{\bf x}||}\leq 1+\delta_k$. The RIP of the measurement matrix ${\bf \Phi}$ guarantees the effective recovery of the $k$-sparse signal ${\bf x} \in {\bf R}^N$ from ${\bf y}={\bf \Phi} \cdot {\bf x}$ (\cite{Candes, CRT}) via linear programming. Though it has been shown that random matrices satisfy the RIP with high probability, in practice the sampling has to be done deterministically. \\

For a $n \times N$ matrix ${\bf \Phi}$ with $N$ columns ${\phi}_1, ....,{\bf \phi}_N$, the coherence is  $\mu_{{\bf \Phi}}=max_{i \neq j} \frac{|<\phi_i,\phi_j>|}{||\phi_i|| \cdot ||\phi_j||}$.  Then $\mu_{{\bf \Phi}} \geq \sqrt{\frac{N}{n(N-n)}}$ from the Welch bound. It is proved in \cite{BDFK,Candes} that a matrix ${\bf \Phi}$ with the coherence $\mu_{{\bf \Phi}}$ satisfies the RIP with the sparsity order $k \leq \frac{1}{\mu_{{\bf \Phi}}}+1$. Thus it is desirable to give explicit construction of matrices with small coherence in compressed sensing.\\

The first systematic deterministic construction of RIP matrices is due to R. DeVore \cite{DeVore}. R. Calderbank and B. Hassibi and their collaborators gave many deterministic constructions of RIP (or statistical isometry property)  matricse in \cite{AHSC,JXC,HCS,CTY}. In 2011 J. Bourgain, S. Dilworth, K.Ford, S. Konyagin and D. Kutzarova \cite{BDFK} gave explicit RIP matrices satisfying $n=o(k^2)$ where $n$ is the row size and $k$ is the sparse order,  from new estimates about exponential sums and additive combinatorics.  There have been many explicit constructions of RIP matrices from various mathematical objects such as chirp sensing codes (\cite{AHSC}), BCH codes (\cite{Amini}), Reed-Muller codes ( \cite{HCS}), orthogonal codes (\cite{YZ}), Reed-Solomon codes (\cite{Mohads}) and expander graphs (\cite{JXC}).  However in these previous constructions, the sizes of the RIP matrices are restricted. This is not desirable in the practical application. \\

A binary contant weight $(n, d, w)$ code is a set of vectors in ${\bf F}_2^n$ such that 1) every condword is a vector of Hamming weight $w$ and 2) the Hamming distance $wt({\bf x}-{\bf y})$ of any two codewords ${\bf x}$ and ${\bf y}$ is at least $d$. Binary constant weight codes have important applications in various information processing problems (\cite{CHT,Chung,Massey,Immink}). To determine the maximal size $A(n, d, w)$ of a $(n, d, w)$ code is a classicla problem in coding theory and has been studied by many authors (\cite{EtzionVardy, Semakov,Graham,table}). For small $n, d, w$, we refer to \cite{table} for the present best known lower bounds for $A(n, d, w)$. The following Gilbert type bound (\cite{Graham}) and Graham-Sloane lower bound are  the most known lower bounds for $A(n, 2d, w)$.\\

{\bf Theorem 1.1 (Gilbert type lower bound ).} {\em $A(n, 2d, w) \geq \frac{\displaystyle{n \choose w}}{\Sigma_{i=0}^{d-1} \displaystyle{w \choose i} \cdot {n-w \choose i}}$.}\\

{\bf Theorem 1.2 (Graham-Sloane bound).} {\em Let $q$ be the smallest prime power satisfying $q \geq n$ then $A(n, 2d, w) \geq \frac{1}{q^{d-1}} \displaystyle{n \choose w}$.}\\

However the binary constant weight codes in Gilbert type lower bound and Graham-Sloane lower bound were not explicitly given. One has to search at least $q^d$ such codes to find the desired one in the Graham-Sloane lower bound (see \cite{Graham}, page 38).\\

A ternary constant weight $(n, d, w)$ code is a set of vectors ${\bf F}_3^n$ such that 1) every condword is a vector of Hamming weight $w$ and 2) the Hamming distance $wt({\bf x}-{\bf y})$ of any two codewords ${\bf x}$ and ${\bf y}$ is at least $d$. The maximal size  $A_3(n, d, w)$ of a $(n, d, w)$ ternary constant weight code have the following Gilbert lower bound (\cite{CheeLing,Svan,Svan1}).\\

{\bf Theorem 1.3 (Gilbert type lower bound)} {\em $A_3(n, d, w) \geq \frac{\displaystyle{n \choose w} \cdot 2^w}{S_{d-1}^{n,w}}$ where $S_{d-1}^{n, w} =\Sigma_{i=0}^{d-1} \Sigma_{j=0}^{min\{[i/2],n-w\}} \displaystyle{w \choose j} \displaystyle{n-w \choose j} \displaystyle{w-j \choose i-2j} 2^j$.}\\

In this paper we give a general constructions of  explicit RIP matrices  with small coherence from general binary and ternary constant weight codes. The resulted RIP matrices will be compared with RIP matrices from pervious constructions. A general  analysis will be given in section 4.\\

\section{Construction}

In this section we prove the following result.\\

{\bf Theorem 2.1.} {\em We have a $n \times A(n, d, w)$ RIP matrix with the coherence $\mu \leq 1-\frac{d}{2w}$.}\\

{\bf Proof.} For a size $A(n, d, w)$ binary constant weight code, we just use its codewords as the column vectors of a $n \times A(n, d, w)$ matrix. Since the intersection of the supports of any two columns has at most $w-\frac{d}{2}$ positions, the conclusion follows directly.\\

{\bf Corollary  2.1.} {\em We have a $n \times A(n, 2w-2, 2)$ RIP matrix with coherence $\mu \leq \frac{1}{w}$.}\\

{\bf $\pm1$ randomized constant-weight-code-based RIP matrices. } \\

If we use $\pm1$'s in the supports of the codewirds of a binary constant weight codes to construct RIP matrices, it is obvious the RIP matrices would have the same size and the coherence of these $\pm1$-randomized RIP matrice are still upper bounded by $1-\frac{d}{2w}$.\\

 In the following construction we take the identification ${\bf F}_3=\{0,1,-1\}$. Then a codeword in a ternary constant weight code can be identified with a vector in ${\bf R}^3$ with coordinates $0, 1, -1$.\\

{\bf Theorem 2.2.} {\em  We have a $n \times A_3(n, d, w)$ RIP matrix with the coherence $\mu \leq 1-\frac{d}{2w}$.}\\

{\bf Proof.}  For any two codewords ${\bf c}_1$ and ${\bf c}_2$ in a ternary constant weight $(n, d, w)$ code we have $d=2w-2|supp({\bf c}_1) \cap supp({\bf c}_2)|+D$ where $D$ is the number of positions in $supp({\bf c}_1) \cap supp({\bf c}_2)$ where the products of the coordinates of ${\bf c}_1$ and ${\bf c}_2$ are $-1$. Then $<{\bf c}_1, {\bf c}_2>=|supp({\bf c}_1) \cap supp({\bf c}_2)|-2D=w-\frac{d}{2}-\frac{3D}{2} \leq w-\frac{d}{2}$. It is clear $||{\bf c}||^2=w$ for each codeword in this ternary constant weight code. The coclusion follows directly.\\

\section{Examples from some best known constant weight codes}

The basic construction of \cite{DeVore} is as follows. For a polynomial $f$ with degree less than or equal to  $r-1$ in ${\bf F}_p[x]$ where ${\bf F}_p$ is a finite field with $p$ elements (here $p$ is a prime number, it can also be used for finite field with $q=p^t$ elements), the length $p^2$ vector $v_f=(f_{(a,b)})$ is determined by its $p^2$ coordinates $f_{(a,b)}$ for $(a, b) \in {\bf F}_p \times {\bf F}_p$. Here $f_{(a,b)}=0$ if $f(a) \neq b$, $f_{(a,b)}=1$ if $f(a)=b$. Then the $p^r$ columns of these  lenght $p^2$ vectors give a $p^2 \times p^r$ matrix. It was proved in [4] that the coherence of this $p^2 \times p^r$ matrix ${\bf D}$ satisfies that $\mu_{{\bf D}} \leq \frac{r-1}{p}$. In the trivial case $r=2$ the RIP matrix from DeVore construction is of size $p^2 \times p^2$ and $\mu=\frac{1}{p}$ for these prime powers.\\

It is well-known $A(q^2, 2(q-1), q)=q^2+q$, we have $q^2 \times (q^2+q)$ RIP matrix with the coherence $\frac{1}{q}$. This is better than the RIP matrices from DeVore construction. However $q^2 \times q^3$ RIP matrix with the coherence $\frac{1}{q}$ was constructed in \cite{Mohads} for each prime power $q$.\\

For the best known binary constant weight codes with small parameters $n, d, w$ we refer \cite{table}. From Theorem 2.1 we list some explicit RIP matrices from constant weight codes.\\

\newpage

\begin{center}
{\bf Table } Explicit RIP matrices from constant weight codes\\
\bigskip
\begin{tabular}{||c|c||}\hline
RIP matrices $\mu$, $n, m$& constant weight codes \\ \hline
$\mu(\Phi) \leq \frac{1}{6}, 40, 45$& $A(40, 10, 6) \geq 45$\\ \hline
$\mu(\Phi) \leq \frac{1}{6}, 42, 55$&$A(42, 10, 6) \geq 55$\\ \hline
$\mu(\Phi) \leq \frac{1}{6}, 45, 57$&$A(45, 10, 6) \geq 57$\\ \hline
$\mu(\Phi) \leq \frac{1}{6}, 47, 63$&$A(47, 10, 6) \geq 63$\\ \hline
$\mu(\Phi) \leq \frac{1}{6}, 50, 72$&$A(50, 10, 6) \geq 72$\\ \hline
$\mu(\Phi) \leq \frac{1}{6}, 51, 76$&$A(51, 10, 6) \geq 76$\\ \hline
$\mu(\Phi) \leq \frac{1}{6}, 55, 87$&$A(55, 10, 6) \geq 87$\\ \hline
$\mu(\Phi) \leq \frac{1}{6}, 56, 91$&$A(56, 10, 6) \geq 91$\\ \hline
$\mu(\Phi) \leq \frac{1}{6}, 61, 111$&$A(61, 10, 6) \geq 111$\\ \hline
$\mu(\Phi) \leq \frac{1}{6}, 63, 126$&$A(63, 10, 6) \geq 126$\\ \hline
$\mu(\Phi) \leq \frac{1}{7}, 49, 56$&$A(49, 12, 7) \geq 56$\\ \hline
$\mu(\Phi) \leq \frac{1}{7}, 55, 63$&$A(55, 12, 7) \geq 63$\\ \hline
$\mu(\Phi) \leq \frac{1}{7}, 56, 71$&$A(56, 12, 7) \geq 71$\\ \hline
$\mu(\Phi) \leq \frac{1}{7}, 61, 72$&$A(61, 12, 7) \geq 72$\\ \hline
$\mu(\Phi) \leq \frac{1}{7}, 62, 79$&$A(62, 12, 7) \geq 79$\\ \hline
$\mu(\Phi) \leq \frac{1}{7}, 64, 88$&$A(64, 12, 7) \geq 88$\\ \hline
$\mu(\Phi) \leq \frac{1}{8}, 64, 72$&$A(64, 14, 8) \geq 72$\\ \hline
$\mu(\Phi) \leq \frac{1}{8}, 71, 80$&$A(71, 14, 8) \geq 80$\\ \hline
$\mu(\Phi) \leq \frac{1}{8}, 72, 89$&$A(72, 14, 8) \geq 89$\\ \hline
$\mu(\Phi) \leq \frac{1}{9}, 81, 90$&$A(81, 16, 9) \geq 90$\\ \hline
\end{tabular}
\end{center}

\subsection{Subspace codes}

A constant dimensional subspace $[n, d, k]_q$ code is a family of $k$ dimensional subspace in ${\bf F}_q^n$ such that the distance $d(U,V)=2k-2dim(U \cap V) \geq d$ satisfied. In a  recent paper \cite{EtzionVardy} of T. Etzion and A. Vardy constructed binary constant weight codes from constant dimension subspace codes. From their result $A(2^{2m-1}, 2^{m+1}-4, 2^m) = 2^{2m-1}+2^{m-1}$ is proved. Thus we have a $2^{2m-1} \times (2^{2m-1}+2^{m-1})$ RIP matrix with the coherence $\mu \leq \frac{1}{2^{m-1}}$. From Theorem 2 in this paper we have a $ q^{tk} \times q^{(t-1)k} \cdot \frac{q^{tk}-1}{q^k-1}$ RIP matrix with the coherence $\mu \leq \frac{1}{q^k}$ (page 3 of the paper). In genral we get the follow construction of RIP matrices  from constant dimension subspace codes.\\

{\bf Theorem 3.1.} {\em If $C$ is an $[n, d = 2t, k]_q$ constant dimension code then there exist a ($q^n-1, 2q^k-2q^{k-t}, q^k-1)$ binary  constant weight code of size $|C|$ and a $(q^n-1,2q^k-2q^{k-t}, q^k)$ binary constant weight code of size $(q^{n-k-1})|C|$. Thues we have $(q^n-1) \times |C|$ RIP matrix with the coherence $\mu \leq \frac{q^{k-t}-1}{q^k-1}$ and $(q^n-1) \times (q^{n-k-1})|C|$ RIP matrix with the coherence $\mu \leq \frac{1}{q^t}$.}\\

\subsection{Steiner systems}

A Steiner system $S(t, w, n)$ is a collection ${\bf B}$ of $w$-subsets taken from an $n$-set $V$  such that each $t$-subset of $V$ is contained in exactly one
element of ${\bf B}$. The close relation of Steiner systems with binary constant weight codes was indicated in \cite{Semakov}. However the existence of the Steiner system for $t \geq 6$ is a difficult and long-standing problem in mathematics (\cite{Wilson}).\\ 

{\bf Corollary 3.1.} {\em If there exists a Steiner system $S(t, w, n)$, we have a $n \times \frac{(n(n-1) \cdots (n-t+1)}{w(w-1) \cdots (w-t+1)}$ RIP matrix with the coherence $\mu \leq \frac{t-1}{w}$. There exists a constant $n_0(w)$ for any given positive integer $w$ such that for any  $n > n_0(w)$ satisfying $n-1 \equiv 0$ $mod$ $w-1$ and $n(n-1) \equiv 0$ $mod$ $w(w-1)$, we have a $n \times \frac{n(n-1)}{w(w-1)}$ RIP matrix with the coherence $\mu \leq \frac{1}{w}$.}\\

{\bf Proof.} From a Steiner system $S(t, w, n)$ a binary constant weight $(n, 2(w-t+1), w)$ can be constructed (\cite{Semakov}). On the other hand we use the famous asmptotical existence result of R. M. Wilson (\cite{Wilson}) to get the sencond conslusion.\\

\section{General analysis}

One of the advantage of constant-weight-code-based RIP matrice is that their sizes can be arbitray. \\

{\bf Theorem 4.1.} {\em For any number $n$ of budged rows and any given sparse order $k$ we can recover the $k$-sparse signals in $N=[\frac{\displaystyle{n \choose kt}}{\Sigma_{i=0}^{(k-1)t-1} \displaystyle{kt \choose i} \cdot {n-kt \choose i}}]$ dimensional real space ${\bf R}^N$ by using the constant-weight-code-based RIP matrix with the coherence $\mu \leq \frac{1}{k}$. Here $t$ can be arbitrary positive integer.}\\

The following result follows from the Graham-Sloane lower bound and Theorem 2.1 directly.\\

{\bf Theorem 4.2.} {\em For any number $n$ of budged rows and any given sparse order $k$ we can recover the $k$-sparse signals in $N=[\frac{\displaystyle{n \choose kt}} {n^{(k-1)t-1}}]$ dimensional real space ${\bf R}^N$ by using the constant-weight-code-based RIP matrix with the coherence $\mu \leq \frac{1}{k}$. Here $t$ can be arbitrary positive integer.}\\

Similarly we have the following result from the Gilbert type lower bound for ternary constant weight codes.\\

{\bf Theorem 4.3.} {\em For any number $n$ of budged rows and any given sparse order $k$ we can recover the $k$-sparse signals in $N=[\frac{\displaystyle{n \choose kt} \cdot 2^{kt}}{S_{2(k-1)t-1}^{n,kt}}]$,  where $S_{2(k-1)t-1}^{n, kt} =\Sigma_{i=0}^{2(k-1)t-1} \Sigma_{j=0}^{min\{[i/2],n-kt\}} \displaystyle{kt \choose j} \displaystyle{n-kt \choose j} \displaystyle{kt-j \choose i-2j} 2^j$, dimensional real space ${\bf R}^N$ by using the ternary-constant-weight-code-based RIP matrix with the coherence $\mu \leq \frac{1}{k}$. Here $t$ can be arbitrary positive integer.}\\

\section{Summary}

Explicit RIP measurement matrices are needed in practical application of compressed sensing to signal processing. In this paper general method of contructing explicit RIP matrices from binary and ternary  constant weight codes is presented.  Thus explicit RIP matrices in compresed sensing can be constructed from constant subspace codes and Steiner systems. The constant-weight-code-basd RIP measurement matrices can be used to sample arbitrary sparse-order signals with arbitrary number of budged rows in some Euclid spaces with suitable dimensions. The RIP matrice from small binary constant weight codes seem to be useful in practice.\\

\bibliographystyle{amsplain}

\end{document}